\documentclass[12pt]{article}
\usepackage{latexsym}
\usepackage{epsfig,amssymb,euscript}
\usepackage{amsmath}
\numberwithin{equation}{section}  

\newsavebox{\ns}
\newsavebox{\dbrane}
\newsavebox{\dbshort}

\def\be{\begin{equation}}
\def\ee{\end{equation}}
\def\bea{\begin{eqnarray}}
\def\eea{\end{eqnarray}}

\def\Dslash{\,\,{\raise.15ex\hbox{/}\mkern-12mu D}}
\def\Dbarslash{\,\,{\raise.15ex\hbox{/}\mkern-12mu {\bar D}}}
\def\delslash{\,\,{\raise.15ex\hbox{/}\mkern-9mu \partial}}
\def\delbarslash{\,\,{\raise.15ex\hbox{/}\mkern-9mu {\bar\partial}}}
\def\pslash{\,\,{\raise.15ex\hbox{/}\mkern-9mu p}}
\def\calDslash{\,\,{\raise.15ex\hbox{/}\mkern-12mu {\cal D}}}

\newcommand\R{\mathbb{R}}
\newcommand\Z{\mathbb{Z}}

\newcommand\C{\mathbb{C}}
\newcommand\T{\mathbb{T}}
\newcommand\dd{\mathrm{d}}
\newcommand{\de}{\partial}

\newcommand{\vol}{\mathrm{vol}}
\newcommand{\diff}{\mathrm{d}}
\newcommand{\e}{\mathrm{e}}
\newcommand{\1}{\mathbf{1}}

\begin{document}
\begin{titlepage}
\begin{center}
\today
{\small\hfill hep-th/0608060}\\
{\small\hfill CERN-PH-TH/2006-160}\\
{\small\hfill HUTP-06/A0031}\\

\vskip 2.5 cm

{\Large\bf  Dual Giant Gravitons 
\vskip 5mm

in Sasaki--Einstein Backgrounds }

\vskip 15mm
{Dario Martelli$^{1}$ and James Sparks$^{2,3}$}\\
\vskip 1 cm

1: Department of Physics, CERN Theory Unit\\
1211 Geneva 23, Switzerland\\
\vskip 0.5cm
2: Department of Mathematics, Harvard University \\
One Oxford Street, Cambridge, MA 02138, U.S.A.\\
\vskip 0.5cm
3: Jefferson Physical Laboratory, Harvard University \\
Cambridge, MA 02138, U.S.A.\\

\vskip 1.5cm

\end{center}

\begin{abstract}
\noindent
We study the dynamics of a BPS D3--brane wrapped on a three--sphere in 
AdS$_5\times L$, a so--called dual giant graviton, 
where $L$ is a Sasakian five--manifold. The phase space 
 of these configurations is the symplectic cone $X$ over $L$, and geometric quantisation naturally
produces a Hilbert space 
of $L^2$--normalisable holomorphic functions on $X$, whose 
states are dual to scalar  chiral BPS operators in the dual superconformal field theory. 
We define classical and quantum partition functions and relate them to 
earlier mathematical constructions by the authors and S.--T. Yau, hep-th/0603021. 
In particular, a Sasaki--Einstein metric
then minimises an entropy function associated with the D3--brane. 
Finally, we introduce a
grand canonical partition function that counts multiple dual
giant gravitons. This is related simply to the index--character of 
the above reference,
and provides a method for counting  multi--trace
scalar BPS operators in the dual superconformal field theory.
\end{abstract}

\end{titlepage}
\pagestyle{plain}
\setcounter{page}{1}
\newcounter{bean}
\baselineskip18pt

\tableofcontents


\section{Introduction and summary}

Recently there has been some interest in counting certain BPS states 
in type IIB string theory on AdS$_5\times S^5$ 
\cite{Romelsberger:2005eg,Kinney:2005ej,Shiraz,giants}. 
In particular, there are two 
classes of classical BPS configurations known as 
giant gravitons and dual giant gravitons, respectively. The former 
consist of D3--branes wrapping three--dimensional supersymmetric 
submanifolds of 
$S^5$, whereas the latter consist of D3--branes wrapping a 
three--sphere in AdS$_5$, and are effectively described by BPS 
point particles in $S^5$. These two sets of classical configurations 
have recently been quantised in \cite{Shiraz} and \cite{giants}, 
respectively. Interestingly, the result is the same in each case, with 
the quantum system being effectively described by a three--dimensional 
harmonic oscillator. 
The AdS/CFT correspondence \cite{Maldacena} in particular relates BPS configurations 
of string theory/supergravity to 
BPS operators of the dual superconformal field theory. 
One may introduce appropriate partition 
functions that count the quantum states of the (dual) giant gravitons 
above, and compare to the counting of 
1/8--BPS scalar chiral primary operators of
${\cal N}=4$ super Yang--Mills theory. 
In  \cite{Shiraz} and \cite{giants} these two calculations were successfully 
matched, both for the giant gravitons and the dual giant gravitons, 
respectively.

As is well--known, the AdS/CFT correspondence 
extends to more general type IIB string theory backgrounds of the 
form AdS$_5\times L$, where $L$ is a Sasaki--Einstein 5--manifold \cite{Kehagias, KW, acharya, MP}. 
Significant progress has been made in understanding these geometries, 
and their $\mathcal{N}=1$ superconformal field theory duals, over the last two years, starting with 
the construction of an infinite family of Sasaki--Einstein 5--manifolds \cite{paper1, paper2}, 
together their dual field theories \cite{toricpaper, quiverpaper}. 

In this paper we shall extend the work of \cite{giants} to 
an arbitrary Sasakian manifold $L$, in particular
quantising the space of BPS dual giant gravitons. We show that the phase space 
of a single BPS dual giant graviton
is precisely the cone $X$ over the Sasakian manifold $(L,g_L)$, equipped 
with the natural symplectic form $\omega$ induced from the contact structure 
on $L$:
\bea\label{symplecticform}
\omega = \frac{1}{2}\diff (r^2\eta)~.\eea
Here $r$ is the conical direction and $\eta$ is the contact one--form on $L$. 
Thus the cone $X$, minus its singular point, is $X_0=\R_+\times L$ where $r$ may be 
thought of as a coordinate on $\R_+$. 
An interesting point about this calculation is 
that the coordinate $r$ was initially, up to a proportionality factor, the radial coordinate $R$ in 
AdS$_5$. Specifically, the two are related by
\bea
r^2 = \frac{2NR^2}{l^2}\eea
where $N$ is the 
number of background D3--branes and $l$ is the AdS$_5$ radius. Recall that the latter is given
by the AdS/CFT formula
\bea
l^4 = 4\pi g N \alpha'^2\eea
with $g$ the string coupling constant. The apex $r=0$ of the cone $X$ corresponds here to a 
D3--brane wrapping a zero--volume three--sphere in AdS$_5$.

The Hamiltonian for the BPS D3--brane is given by
\bea
H_\mathrm{BPS} & = & \frac{r^2}{2l}~.
\eea
Recall that every K\"ahler cone is equipped with a holomorphic 
Killing vector field
\bea
\xi = \mathcal{J}\left(r\frac{\partial}{\partial r}\right)\eea
where $\mathcal{J}$ is the complex structure tensor on $X$. 
The BPS D3--brane Hamiltonian is then precisely the Hamiltonian function for the 
vector field $l^{-1}\xi$.

When the Sasakian manifold $(L,g_L)$ is toric, 
the isometry group contains  $U(1)^3$ by 
definition, 
and there are correspondingly 
three conserved momenta $P_{\phi_i}$, $i=1,2,3$, in the D3--brane dynamics.
These are the momenta conjugate to the cyclic coordinates $\phi_i$ parametrising the three--torus 
$\T^3=U(1)^3$. The Hamiltonian for BPS configurations may then be written as
\bea
H_\mathrm{BPS} & = & \frac{1}{l}b_i P_{\phi_i}~,
\eea
where $b_i$ are the components of the Reeb vector field in the above basis 
\bea
\xi = b_i \frac{\de}{\de \phi_i}~.
\eea

Equipped with a classical phase space $(X,\omega)$, which is also K\"ahler, 
together with a Hamiltonian $H=r^2/2l$, it is straightforward to 
quantise the system using geometric quantisation \cite{Woodhouse}, which in the 
present set--up is also similar to Berezin's quantisation \cite{berezin}. 
Using the complex polarisation induced by the complex structure on $X$,
one finds that the Hilbert space $\mathcal{H}$ is the $L^2$--completion of the space 
of $L^2$--normalisable holomorphic functions on $X$ with inner product
\bea\label{inner}
\langle f_1,f_2\rangle = \int_X {f_1} \bar{f_2} \, \e^{-r^2/2} \frac{\omega^3}{3!}~.\eea
Applying the standard rules of geometric quantisation, 
in particular the Hamiltonian becomes
\bea
l \hat{H} = -i \mathcal{L}_{\xi} 
\eea
acting on $\mathcal{H}$. Thus quantum states of definite energy are precisely holomorphic 
functions of definite charge under the Reeb vector field. In field theory language, 
the Hamiltonian governing the dynamics of BPS states is precisely given by the R--charge,
or equivalently the dilatation, operator.

One may then define, in the usual way, the classical and quantum 
partition functions:
\bea\label{class}
Z_{\mathrm{classical}}(\beta) & = & \int_X \e^{-\beta H}\frac{\omega^3}{3!}
 = \frac{8l^3}{\beta^3} \vol[L]\\
\label{quant}
Z_{\mathrm{quantum}}(\beta) & = & \mathrm{Tr}_{\mathcal{H}} \e^{-\beta\hat{H}}~.\eea
The second equality in (\ref{class}), which is straightforward to 
derive, was essential in \cite{MSY2}. One of the central results of 
the latter reference was that one can localise this expression for the volume, 
with respect to $\xi$, by appropriately resolving $X$ and using  
the formula of Duistermaat and Heckman \cite{DH}. This leads to a formula for 
the volume $\vol[L]$ which is a rational function of $\xi$ with rational coefficients. 
These coefficients are given by certain Chern classes and weights. 
On the other hand, the quantum partition function 
(\ref{quant}) was called the holomorphic partition function in \cite{GMSY} 
and is closely related to the character defined in \cite{MSY2}. 
Indeed, the latter reference implies that $Z_{\mathrm{quantum}}(\beta)$ 
has a pole of order $3$ as $\beta\rightarrow 0$, with 
\bea\label{agree}
Z_{\mathrm{quantum}}(\beta) \sim \frac{1}{(2\pi)^3}
Z_{\mathrm{classical}}(\beta), \qquad \beta\rightarrow 0~.\eea

For much of this paper we formally 
consider an arbitrary Sasakian metric on $L$. The above partition 
functions then also become functions of the Reeb vector field $\xi$ as one varies 
the background metric \cite{MSY2}. However, in order 
to satisfy the type IIB equations of motion, the metric $g_L$ on $L$ must be Einstein 
of positive curvature. One can formally define an entropy function for a
BPS dual giant graviton from its classical partition function (\ref{class}). 
A Sasaki--Einstein background then minimises 
this entropy function with respect to the Reeb vector field $\xi$. This is rather analogous to Sen's entropy function 
for black holes \cite{sen}.

Finally, we consider a grand canonical partition function that counts 
multiple BPS dual giant gravitons. These may effectively be described as 
$n$ indistinguishable particles, with $n$ bounded from above by the 
number of background D3--branes $N$ \cite{ramgoolam, nemani, smith}.
Suppose that the K\"ahler cone $(X,\omega)$ admits a holomorphic $U(1)^s$ isometry, generated by
vector fields $J_i$, $i=1,\ldots,s$. As discussed in section \ref{quantum}, these act as Hermitian operators $\hat{P}_i=-i\mathcal{L}_{J_i}$ on 
$\mathcal{H}$ and we may thus define the grand canonical partition 
function
\bea\label{bored}
{\cal Z}(\zeta,\mathbf{q},X) = \mathrm{Tr}_{\mathcal{H}_{\mathrm{multi}}} 
\zeta^{\hat{N}}\mathbf{q}^{\mathbf{\hat{P}}}\eea
where $\mathcal{H}_{\mathrm{multi}}$ is the multi BPS dual giant 
Hilbert space and $\hat{N}$ 
is the operator that counts the number of dual giant gravitons in a given 
state. In particular, 
the coefficient of $\zeta$ is precisely the character
\bea
C(\mathbf{q},X) = \mathrm{Tr}_{\mathcal{H}} \mathbf{q}^{\mathbf{\hat{P}}}\eea
of \cite{MSY2}. Moreover, setting
\bea
\xi = \sum_{i=1}^s b_i J_i\eea
and $q_i=\exp(-\beta b_i/l)$, the character is precisely the quantum partition 
function (\ref{quant}).

It is straightforward to write the partition function (\ref{bored}) entirely in terms of the 
character:
\bea
{\cal Z} (\zeta,\mathbf{q},X) = \exp\left[ \sum_{n=1}^\infty \frac{\zeta^n}{n}\, C(\mathbf{q}^n,X)\right]~,
\eea
and thus the partition function $Z_N(\mathbf{q},X)$ for $N$ BPS dual giant gravitons may be extracted rather simply 
as the coefficient of $\zeta^N$ in this expression. 
$Z_N(\mathbf{q},X)$ may be interpreted as the trace of the action of $\mathbf{q}\in (\C^*)^s$ on 
the space of holomorphic functions on the symmetric product space
$\mathrm{Sym}^N X$.

In the AdS/CFT dual superconformal field theory, 
this is 
precisely the generating function that
counts mesonic scalar chiral primary operators 
according to their $U(1)^s$ flavour charges. 
Indeed, if a SCFT arises from the IR limit of $N$ D3--branes 
at an isolated singularity $X$, then
the classical vacuum 
moduli space should be the symmetric product $\mathrm{Sym}^N X$. 
The coordinate ring of holomorphic functions on this variety is 
the symmetric product of the coordinate ring of $X$. 
The Hilbert space of $N$ BPS dual giants above is then spanned 
by the same set of generators (as a $\C$--algebra) 
as this ring. Thus our counting of chiral primaries in the CFT, obtained via counting $N$ 
dual giant gravitons
in the geometry, agrees  with the results of \cite{Kinney:2005ej}, based on group--theoretic techniques.

The plan of the rest of the paper is as follows. In section \ref{bigboys} 
we analyse the classical dynamics of a D3--brane probe wrapping 
an $S^3\subset \mathrm{AdS}_5$, focusing in particular on BPS 
configurations. In section \ref{quantum} we quantise the corresponding 
phase space using geometric quantisation. In section \ref{partition} we 
write down the classical and quantum partition functions, in particular 
relating them to the results of \cite{MSY2}. The volume minimisation 
of the latter reference is related to minimising an entropy function for the D3--brane. 
Finally, in section \ref{grand}, we study a grand canonical partition 
function that counts multiple BPS dual giant gravitons, and discuss  
the relation to counting BPS chiral primary operators in the AdS/CFT dual 
superconformal field theory.


\section{Dual giant gravitons}
\label{bigboys}

In this section we study the dynamics of a
dual giant graviton in AdS$_5\times L$, where $(L,g_L)$ is 
an arbitrary Sasakian 5--manifold. In order that this background satisfies 
the type IIB supergravity equations one requires $g_L$ to be a positively 
curved Einstein metric, but for the most part this will be inessential in 
what follows -- the important feature is the Sasakian structure. 
The dynamics essentially reduces to that of a point 
particle on $L$, and the BPS configurations are described by  
BPS geodesics on $L$. The BPS phase space is precisely the cone $(X,\omega)$ 
based on $L$, equipped with the standard symplectic form $\omega$ induced
 from the contact structure on $L$. Moreover,
the Hamiltonian restricted to these configurations
 is proportional to the Hamiltonian function for the Reeb
vector field $\xi$ on $X$.

\subsection{Hamiltonian dynamics and phase space}
\label{ggiants}

We begin with the direct product metric on AdS$_5\times L$
\bea
\diff s^2 = g_{MN}\diff X^M\diff X^N = \diff s^2_{\mathrm{AdS}_5}+l^2\diff s^2_L\eea
where $X^M$, $M=0,\ldots,9$, 
are local coordinates on AdS$_5\times L$, $\diff s^2_L$ is the line element 
of a Sasakian metric on the 5--manifold $L$, and $l$ is the 
AdS$_5$ radius. 
One may introduce global coordinates on AdS$_5$ with line element 
\bea\diff s^2_{\mathrm{AdS}_5} = - V(R)\diff t^2 + \frac{1}{V(R)}\diff R^2 + R^2 (\diff\theta^2 + \cos^2 \theta
\diff\phi_1^2 +\sin^2 \theta \diff\phi_2^2)\eea
where
\bea
V(R) & = & 1 + \frac{R^2}{l^2}~.\\
\eea
Here $t$ is the usual global time on AdS$_5$. The coordinate $R$ is then a radial 
coordinate on the constant time sections, foliating the latter with round three--spheres.

The simplest way of defining a Sasakian manifold $(L,g_L)$ is to say that the corresponding 
metric cone $(X,g_X)$, with line element
\bea\label{bottom}
\diff s^2_X = \diff r^2 + r^2 \diff s^2_L~,\eea
is K\"ahler, although there exist other, more intrinsic, definitions. 
An important fact in what follows is that any Sasakian metric may be written locally as
\bea
\diff s^2_L = h_{\alpha\beta} \diff x^{\alpha} \diff x^{\beta} + (\diff\psi +\sigma)^2~.\eea
Here the Reeb vector field is
\bea
\xi = \mathcal{J}\left(r\frac{\partial}{\partial r}\right) = \frac{\partial}{\partial\psi}\eea
which has norm one on the link $\{r=1\}$, which is a copy of $L$. The metric transverse to the orbits of $\xi$ is given locally in components
by $h_{\alpha\beta}(x)$, $\alpha,\beta=1,\ldots,4$, and is also a K\"ahler metric.
The contact one--form, metrically dual to $\xi$, is, in these local coordinates,
\bea
\eta = \diff\psi + \sigma\eea
and satisfies
\bea
\diff\eta = \diff\sigma = 2\omega_T,\eea
where $\omega_T$ is the transverse K\"ahler form. In particular, the K\"ahler 
cone metric on $X$ then has K\"ahler form
\bea
\omega = \frac{1}{2}\diff(r^2\eta) = \frac{1}{2}i\partial\bar{\partial}r^2~.\eea
For further details on 
Sasakian geometry we refer the reader to \cite{MSY2} and references therein.

The dynamics of a D3--brane propagating in this background is described 
by the usual world--volume action, comprising the two terms
\bea
S_{\mathrm{D3}} = S_{\mathrm{DBI}}+ S_{\mathrm{WZ}} = -T_3\int \dd^4 \zeta \sqrt{- \det  G_{\mu\nu} } + T_3 \int C_{4}
\eea
where 
\bea
G_{\mu\nu} = \frac{\de X^M}{\de \zeta^\mu} \frac{\de X^N}{\de \zeta^\nu} g_{MN}
\eea
is the pull--back of the spacetime metric to the D3--brane world--volume, parametrised by coordinates
$\{\zeta^0=\tau,\zeta^1,\zeta^2,\zeta^3\}$, $C_4$ is the (pull--back of the) Ramond--Ramond four--form potential 
of type IIB supergravity, and $T_3$ is the D3--brane tension.

We wish to study configurations in which the D3--brane wraps a round 
three--sphere in AdS$_5$, a so--called dual giant graviton 
\cite{McGreevy:2000cw}. 
We thus choose the following embedding $X^M(\zeta^\mu)$:
\bea\label{embedding}
 &&t=\tau \quad R=R(\tau) \quad \theta = \zeta^1 \quad \phi_1 = \zeta^2 \quad \phi_2 = \zeta^3\nonumber\\
&&\psi = \psi(\tau) \quad x^{\alpha} = x^{\alpha}(\tau)~.
\eea
The self--dual Ramond--Ramond five--form for this background is 
\bea
F_5 & = &-\frac{4}{l}\left( \mathrm{Vol}[\mathrm{AdS}_5] + l^5\mathrm{Vol}[L]\right) 
\eea
and it is then easily verified that
the pull--back of a choice of $C_4$ to the world--volume of the D3--brane, under the embedding (\ref{embedding}), is simply 
\bea
C_4& = & \frac{R^4}{l} \sin\zeta^1 \cos \zeta^1 \diff\tau\wedge\diff\zeta^1\wedge\diff\zeta^2\wedge\diff\zeta^3~.
\eea
A short calculation reveals that the determinant factor reads
\bea
\sqrt{- \det G_{\mu\nu}} & = & R^3 \cos\zeta^1 \sin \zeta^1 \Delta^{1/2}~,
\eea 
where we have defined
\bea
\Delta \equiv V(R)-\frac{\dot R^2}{V(R)} - l^2\left[ h_{\alpha\beta}(x)\dot x^{\alpha} \dot x^{\beta}  
+ (\dot \psi +\sigma_{\alpha} \dot x^{\alpha})^2\right]
\eea
and a dot denotes differentiation with respect to $\tau$. 
Finally, integrating the action over the $S^3$ and using the formula (see {\it e.g.} \cite{golia})
\bea
2\pi^2 T_3 & = & \frac{N}{l^4}
\eea
for the D3--brane tension, we obtain the effective point--particle Lagrangian
\bea
L & = & -\frac{N}{l^4}R^3\left[ \Delta^{1/2} - \frac{R}{l}\right]~.
\label{effect}
\eea

To proceed, it is convenient to pass to the Hamiltonian formalism.
The canonical momenta are
\bea
P_R & \equiv & \frac{\partial L}{\partial \dot{R}} = \frac{N R^3}{l^4V(R)\Delta^{1/2}}\dot R \nonumber\\
P_\psi & \equiv & \frac{\partial L}{\partial \dot{\psi}} = \frac{N R^3}{l^2\Delta^{1/2}} (\dot \psi + \sigma_{\alpha}
\dot x^{\alpha})\nonumber\\
P_{x^{\alpha}} & \equiv & \frac{\partial L}{\partial \dot{x^{\alpha}}} =  \frac{N R^3}{l^2\Delta^{1/2}} \left(h_{\alpha\beta}\dot x^{\beta} +(\dot \psi + \sigma_{\gamma}
\dot x^{\gamma})\sigma_{\alpha}\right)~.
\eea
The Hamiltonian is then 
\bea
H &  = &  
\frac{NR^3}{l^4}\left[V(R)^{1/2}\Omega^{1/2}-\frac{R}{l}\right] ~,
\label{hami}
\eea
where we have defined
\bea
\Omega & = & 1 + \frac{l^6}{N^2R^6}\left[ l^2 V(R) P_R^2 + P^2_\psi +
 h^{\alpha\beta} (P_{x^{\alpha}}- P_\psi \,\sigma_{\alpha}) (P_{x^{\beta}}- P_\psi \,\sigma_{\beta})\right]~,
\eea
and $h^{\alpha\beta}$ denotes the matrix inverse of $h_{\alpha\beta}$. It is a standard exercise to verify that 
the Hamiltonian equations of motion
\bea
\dot P_A = - \frac{\de H}{\de Q^A} \qquad\quad  \dot Q^A = \frac{\de H}{\de P_A}~,
\eea
where $Q^A$ and $P_A$ collectively denote the six coordinates and their conjugate momenta, respectively, 
admit the following solutions  
\bea
 &&\dot R = 0\qquad \dot \psi = \frac{1}{l} \qquad \dot x^{\alpha} = 0 \nonumber\\
&&  P_R = 0 \qquad P_\psi = \frac{NR^2}{l^2} \qquad 
 P_{x^{\alpha}}=\frac{NR^2}{l^2}\sigma_{\alpha}~.
 \label{BPSsolu}
\eea 
In fact, one can show that these solutions are precisely the set of $\kappa$--symmetric, or BPS, solutions. The calculation 
is again standard, although slightly lengthy. The details may 
be found in appendix \ref{BPS}.

The Hamiltonian, restricted to these configurations, is
\bea
H_{\mathrm{BPS}}  = \frac{N}{l^3}R^2 = \frac{1}{l}P_\psi~,
\eea
and in particular is proportional to $R^2$. We define
\bea\label{rescale}
r^2 \equiv  \frac{2NR^2}{l^2}\eea
so that
\bea
H_{\mathrm{BPS}} =  \frac{r^2}{2l}~.
\eea
For $(X,g_X)$ the K\"ahler cone over $(L,g_L)$, with metric (\ref{bottom}),
$r^2/2$ is 
precisely the Hamiltonian function generating 
the flow along the Reeb vector field; that is, 
\bea
\diff (r^2/2) = -\xi\lrcorner\omega~.
\eea
This Hamiltonian function for $\xi$ was essential in reference \cite{MSY2} for computing the volume of the link $(L,g_L)$ 
using the localisation formula of Duistermaat and Heckman.
Here, we recover this Hamiltonian from a different, and more physical, perspective. Indeed, we find that $r^2/2$
here is precisely the momentum canonically conjugate to the Reeb vector field. Interestingly, 
the conical coordinate $r$ on phase space is effectively constructed from the radial coordinate $R$ in AdS$_5$.

To make this correspondence more precise, we turn to analysing the resulting reduced phase space.
Given the constraints (\ref{BPSsolu}), the reduced phase space may be parametrised
by the six local coordinates
\bea
Q^A & = & (R,\psi,x^1,\ldots,x^4)^A~.
\eea
Indeed, it is straightforward to see that this phase space is naturally a copy of the cone $X$ over $L$ with conical 
coordinate $r$. The tip of the cone $r=0$ corresponds to the singular configuration in which the 
D3--brane wraps a zero--volume $S^3$ in AdS$_5$.
A standard way to obtain the symplectic structure on phase space is to compute the Dirac brackets.
Let us repackage the six constraints as follows:
\bea
f_1 =  P_R = 0 \qquad f_2 = P_\psi -\frac{N}{l^2}R^2 =0 \qquad f_{\alpha+2} = P_{x^{\alpha}}-
\frac{N}{l^2}R^2\sigma_{\alpha} =0
\eea
and compute the following Poisson brackets
\bea
\{Q^A,Q^B \}_{PB} = 0 \qquad \{Q^A,f_B\}_{PB}= \delta^{A}_{B}
\eea
and
\bea
M_{AB}  = \{f_A,f_B \}_{PB}= \frac{N}{l^2}2R\left(
\begin{array}{ccc}
0 & 1 &  \,\sigma_{\alpha}   \\
-1 & 0 & \, 0  \\
-\sigma_{\beta} & 0 &  \, R(\omega_T)_{\alpha\beta}  \\
\end{array} 
\right)~.
\eea
We thus obtain
\bea
\{Q^A,Q^B\}_{DB}  =  \{Q^A,Q^B\}_{PB}- \{Q^A,f_C\}_{PB} M_{CD}^{-1}\{f_D,Q^B\}_{PB}=
 M^{AB}
\eea
where $M^{AB}$ is the matrix inverse of $M_{AB}$, 
giving the symplectic structure 
\bea
\omega  =  \frac{N}{l^2} \dd (R^2 \eta ) = \frac{1}{2}\diff(r^2\eta)~. 
\label{symplcone}
 \eea
Thus we recover the standard symplectic structure on the K\"ahler cone $(X,g_X)$. 
Alternatively, the same result may be obtained by computing
the symplectic one--form $\nu$ on phase space: 
\bea
\nu = P_R \diff R + P_{\psi} \diff \psi + P_{x^{\alpha}}\diff x^{\alpha} = \frac{N}{l^2} R^2\eta = \frac{1}{2}r^2\eta~.\eea

\subsection{Toric geometries}
\label{toricsub}

In this subsection we discuss the particular case that $(L,g_L)$ is a toric 
Sasakian manifold. This is of course a subcase of the general discussion in the previous subsection, but it is
nevertheless instructive to analyse explicitly. 
When  $(L,g_L)$ is toric, by definition there is  at least a $U(1)^3$ isometry group, 
and there are correspondingly 
three conserved momenta in the D3--brane dynamics considered in the previous subsection. 
These are dual to certain global flavour charges in the AdS/CFT dual conformal field theory, 
with the R--charge arising as a linear combination of these three charges.

The group $\T^3=U(1)^3$  acts Hamiltonianly on the K\"ahler cone $(X,g_X)$, which in the present case means 
that the $\T^3$ action preserves the K\"ahler form $\omega$. A general toric K\"aher cone $(X,g_X)$ metric may be written as \cite{MSY}
\bea
\dd s^2_X & = & G_{ij}\dd y^i \dd y^j + G^{ij}\dd\phi_i \dd\phi_j~.
\label{toricmetric}
\eea
Here $\phi_i\sim \phi_i +2\pi$ are angular coordinates on $\T^3$, with corresponding moment map coordinates
$y^i$, $i,j=1,2,3$. The $y^i$ are homogeneous solutions to
\bea
\diff y^i = -\frac{\partial}{\partial\phi_i}\lrcorner \omega~.\eea
The symplectic form $\omega$ is then simply
\bea
\omega = \diff y^i\wedge\diff\phi_i\eea
while the metric is parametrised by the Hessian matrix $G_{ij}$ of the symplectic potential $G(\mathbf{y})$
\bea
G_{ij} = \frac{\de^2 G}{\de y^i \de y^j}~,
\eea
with $G^{ij}$ denoting the matrix inverse of $G_{ij}$. The map
\bea\label{momentmap}
\mu &:& X \rightarrow\R^3\nonumber\\
    && (y^i,\phi_i)\mapsto y^i\eea
is called the moment map, and the image $\mu(X)=\mathcal{C}^*$ is a strictly convex rational polyhedral cone
\bea
\mathcal{C}^* = \left\{ \mathbf{y} \in \R^3 ~|~ l^a(\mathbf{y}) = v^a_i y^i \geq 0 \, , a=1,\dots, d \right\}~.
\eea
Thus it is the convex cone formed by $d$ planes through the origin of $\R^3$. The inward--pointing normal vectors to 
these planes may be taken to be integral primitive vectors $v^a\in\Z^3$, $a=1,\ldots,d$. The symplectic potential 
$G(\mathbf{y})$ is then a function on $\mathcal{C}^*$ with a certain prescribed singular behaviour near the bounding 
planes. Moreover, $G_{ij}$ is required to be homogeneous degree minus one in the $y^i$.
For further details, the reader is referred to \cite{MSY}.

The metric $\dd s^2_L$ on the link is simply given by the restriction of the metric (\ref{toricmetric})
to the hypersurface
\bea
2 b_i y^i & = & 1~.
\label{link}
\eea
Here the constants $b_i$ are the components of the Reeb vector field in the above basis:
\bea
\xi = b_i\frac{\partial}{\partial\phi_i}~.\eea

To study the dynamics of a D3--brane wrapped on an $S^3$ inside AdS$_5$, one proceeds as before.
 The Hamiltonian can be written as in (\ref{hami}), with 
\bea
\Omega & = & 1 + \frac{l^6}{N^2R^6}\left[ l^2 V(R)P^2_R  + G^{ij} P_{y^i} P_{y^j} +
 G_{ij}P_{\phi_i} P_{\phi_j} \right]~.
\eea
where the canonical momenta are
\bea
P_{y^i} = G_{ij} \dot y^j \qquad \quad 
P_{\phi_i} =  G^{ij} \dot \phi_j~.
\label{toricmoments}
\eea
The $\phi_i$ are cyclic coordinates, so that it immediately follows that the $P_{\phi_i}$
are constant. 
The BPS solutions are given by
\bea
&& \dot R = 0 \qquad\dot y^i = 0\qquad  \dot \phi_i = \frac{1}{l} b_i\nonumber\\
&& P_R = 0 \qquad  P_{y^i} = 0 \qquad  P_{\phi_i} = \frac{2NR^2}{l^2}y^i 
\label{toricgiants}
\eea
which also solve the Hamiltonian equations. 
Note that $y^i$ are not independent variables. Thus one might   
introduce a Lagrange multiplier and implement the usual 
Hamiltonian description of constrained systems. 
Alternatively, one can simply solve for one variable in terms of the other two. 
For instance, one can choose $y^2,y^3$ as independent variables, and regard  $y^1$ as a function of these.
Taking note of this, and using formulae from \cite{MSY}, one can indeed verify that (\ref{toricgiants}) 
is a solution. 
We define
\bea
r^i \equiv P_{\phi_i} = \frac{2NR^2}{l^2}y^i = r^2y^i~,
\eea
so that the $r^i$ effectively become moment map coordinates on the cone $X$, 
and the reduced phase  space is parametrised by the coordinates
$(r^i,\phi_i)$ and  has symplectic form
\bea
\omega = \diff r^i\wedge\diff\phi_i~.
\eea
Finally, the BPS Hamiltonian is
\bea
\label{toricBPS}
H_\mathrm{BPS} = \frac{1}{l} \, b_i P_{\phi_i}~.
\eea
Notice that we could have arrived at the same results by simply implementing the change of coordinates
discussed in appendix \ref{change}.

\subsubsection*{Example: $1/8$--BPS dual giants in AdS$_5\times S^5$}

As an example of this formalism, let us recover the results of \cite{giants}. Of course, $1/8$--BPS configurations
in an $S^5$ background preserve the same supersymmetry as $1/2$--BPS configuration in a Sasaki--Einstein background.
Equivalently, $1/8$--BPS operators of ${\cal N}=4$ super Yang--Mills preserve the same supersymmetry as $1/2$--BPS operators 
in an ${\cal N}=1$ SCFT. 

 We view $S^5$ as a toric Sasaki--Einstein manifold, with $U(1)^3\subset SO(6)$ 
the Cartan subgroup. Setting 
$y^i = (\mu^i)^2/2$
and taking 
\bea
G(\mathbf{y})& = & \frac{1}{2}\sum_{i=1}^3 y^i \log y^i
\label{Gsphere}
\eea
the toric metric (\ref{toricmetric}) reads
\bea
\dd s^2_{\C^3} = \sum_{i=1}^3\left( (\dd\mu^i)^2 + (\mu^i)^2 \dd\phi_i^2\right)
\eea
while the constraint is
\bea
\sum_{i=1}^3 (\mu^i)^2 = 1~.
\label{constr}
\eea
If one wishes, one may introduce unconstrained angular variables to solve for (\ref{constr}).
It is straightforward to check\footnote{Notice that here the toric fan is 
generated by the
standard orthonormal basis for $\R^3$, $v^i_a=\delta^i_a$.}, 
for example using formulae in \cite{MSY}, that the Reeb vector field is 
$b_i = (1,1,1)$.
Inserting this into the general toric forula (\ref{toricBPS}) for the BPS Hamiltonian 
one obtains 
\bea
H_{S^5} = \frac{1}{l} (P_{\phi_1}+P_{\phi_2}+P_{\phi_3})~,
\eea
which is the result presented in \cite{giants}.

\subsection{Relation to BPS geodesics}

In this subsection, we point out that the dynamics of a BPS point--particle\footnote{BPS geodesics in the
$Y^{p,q}$
\cite{paper1,paper2} and $L^{a,b,c}$ \cite{CLPP,Martelli:2005wy} manifolds were considered in 
\cite{BK1,BK2}, and were argued to be related to  chiral primary operators of the dual quiver gauge theories 
\cite{quiverpaper, Franco:2005sm, Butti, BK2}.} in an
arbitrary Sasakian manifold is equivalent to that of a BPS dual giant graviton, previously discussed. 
Geometric quantisation of this dynamics, to be discussed in section 3, will then lead to a precise relation to chiral 
primary operators in the dual conformal field theory.

We consider the motion of a free point--like particle in the metric
\be
\dd s^2 = -\dd t^2 + \dd s^2_L
\ee
where $(L,g_L)$ is a Sasakian manifold, with Reeb vector field $\xi=\partial/\partial\psi$.
We therefore consider the following action\footnote{We suppress overall multiplicative factors in the action.}
\be
S =\int \dd\tau \left[-\dot t^2 + h_{\alpha\beta}(x)\dot x^{\alpha} \dot x^{\beta}  
+ (\dot \psi +\sigma_{\alpha} \dot x^{\alpha})^2\right]
\ee
where dots denote derivatives with respect to $\tau$, and $\alpha,\beta=1,\ldots,4$. Passing to the Hamiltonian formalism, we have
\be
H = - p_t^2+ p_\psi^2 +h^{\alpha\beta} (p_{\alpha} - p_\psi \sigma_{\alpha})(p_{\beta} -p_\psi \sigma_{\beta} )~,
\ee
in terms of the conjugate momenta.
As $t$ and $\psi$ are cyclic coordinates, their conjugate momenta are constant: 
$p_t=E$, $p_\psi=\lambda$.
Setting to zero the total Hamiltonian, as follows from reparametrisation invariance, 
one obtains an expression for the energy:
\be
E^2 = \lambda^2 +h^{\alpha\beta} (p_{\alpha} - \lambda\sigma_{\alpha})(p_{\beta} -\lambda \sigma_{\beta} )~.
\ee
This is positive definite; in particular we have the bound
\bea
E\geq \lambda~.
\label{BPSbound}
\eea
It is then natural to define BPS geodesics as those trajectories for which 
the inequality (\ref{BPSbound}) is saturated. This immediately implies that 
\bea
p_{\alpha} = \lambda \sigma_{\alpha}~.
\eea
One can check that the full solution to the equations of motion is given by
\bea
\dot x^{\alpha} = 0~,\qquad \dot \psi = \lambda~.
\eea
Thus a BPS geodesic is precisely an orbit of the Reeb vector field $\xi$, with the particle moving at 
constant speed $\lambda$ in this direction. Thus the configuration space is a copy of 
the cone over $(L,g_L)$: for each point in $L$ there is a unique BPS geodesic starting at that 
point\footnote{Assuming $\lambda>0$.}. 
The conical direction may then identified with the speed as
\bea
\lambda = r^2/2~.
\eea
Indeed, one easily checks that the symplectic form on phase space is
\bea
\nu = p_{\psi}\diff\psi+p_{\alpha}\diff x^{\alpha} = \lambda\eta = \frac{1}{2}r^2\eta\eea
with this identification. Moreover, the energy is 
\bea
E = \frac{r^2}{2}
\eea
which is the same as the rescaled BPS Hamiltonian $lH_{\mathrm{BPS}}$ for 
the BPS dual giant gravitons.

Finally, for toric geometries, using the change of coordinates in appendix \ref{change}
it is straightforward to write down the BPS geodesics in terms of the Reeb vector: 
\bea
\dot \phi_i = \lambda \, b_i  \qquad  \quad \dot y^i =0 
\eea
and obtain the following expression for the energy 
\bea
E  =  b_i p_{\phi_i}~.
\eea


\section{Geometric quantisation}
\label{quantum}

In this section we quantise the BPS dual giant gravion. This is a fairly routine 
exercise in applying geometric quantisation to the phase space $X$ derived in the 
previous section. The result is rather simple: the Hilbert space $\mathcal{H}$ is the space of $L^2$--normalisable
holomorphic functions on $X$, with respect to the inner product (\ref{inner}). There is 
then a standard map from quantisable functions on $X$ to operators on $\mathcal{H}$, which in particular 
maps the Hamiltonian $H_{\mathrm{BPS}}=r^2/2l$ to $\hat{H}=-il^{-1}\mathcal{L}_{\xi}$. Thus states of 
definite energy are described by holomorphic functions on $X$ of definite R--charge under the 
Reeb vector field $\xi$.

\subsection{Hilbert space}

Given a phase space $X$, with symplectic form $\omega$, one would like to quantise the classical 
system $(X,\omega)$. Thus, one would like to associate a Hilbert space $\mathcal{H}$ in a natural way to $(X,\omega)$. 
Moreover, to every classical observable, namely a function $A$ on 
$X$, one would like to associate a symmetric operator $\hat{A}$ on $\mathcal{H}$. According to 
Dirac, the map $A\rightarrow \hat{A}$ should be linear and map Poisson brackets to commutators. 
Thus, operators should form a Hilbert space representation of the classical observables.

Given any symplectic manifold $(X,\omega)$, a natural Hilbert space is simply the $L^2$--completion of 
the space of smooth $L^2$--normalisable complex--valued functions on $X$, with norm being 
$\langle f_1,f_2\rangle = \int_X f_1\bar{f_2} \omega^n/n!$. One could then map the function 
$A\mapsto -iX_A$ where $X_A$ is the Hamiltonian vector field for $A$, which by definition satisfies
\bea
\diff A = -X_A\lrcorner \omega~.\eea
In fact this map from functions on $X$ to vector fields is a Lie algebra homomorphism with respect to the Poisson bracket
\bea
[A,B]=\omega^{ij}\partial_iA\partial_jB\eea
and the usual Lie bracket $[X_A,X_B]$ of vector fields.
However, there are various problems with this, not least that the constant function is mapped to 
zero, so that for example position and momentum then commute. Moreover, since our phase space 
$X$ is a non--compact cone, the wavefunctions would have to have very rapid decrease in the conical 
direction.

Geometric quantisation is an attempt to solve this quantisation problem in general\footnote{For 
a review, see reference \cite{Woodhouse}.}. The first step 
is to define an Hermitian line bundle $\mathcal{L}$ over $X$ with unitary connection for which $-2\pi i\omega$ is the curvature 2--form. 
In general, a necessary condition is that the periods of $\omega$ are integral. However, in our case, 
$\omega$ is in fact exact
\bea
\omega = \frac{1}{2}\diff (r^2\eta) = \frac{1}{2}i\partial\bar{\partial} r^2\eea
and moreover is a K\"ahler form with a globally defined K\"ahler potential $r^2/2$. This also happens to be the Hamiltonian for 
the BPS D3--brane of course, up to a factor of the AdS radius $l$. Since $\omega$ is exact, the line bundle $\mathcal{L}$ is trivial 
as a complex line bundle, and we thus take
\bea
\mathcal{L} = \C\times X~.\eea
In particular, sections of $\mathcal{L}$ may be identified with complex--valued functions on $X$. Given two functions $f_1$, $f_2$, viewed 
as sections of $\mathcal{L}$, their pointwise inner product is 
\bea
(f_1,f_2) = f_1\bar{f_2}\exp(-r^2/2)~.\eea
This metric has the property that it is compatible with the connection
\bea
\nabla = \diff - i\nu\eea
where
\bea\label{oneform}
\nu = -\frac{1}{2}i\partial r^2\eea
is a connection 1--form with $\diff\nu = \omega$. Compatability means that
\bea
\diff (f_1,f_2) = (\nabla f_1, f_2) + (f_1,\nabla f_2)\eea
which the reader may easily verify. 

Thus, a first attempt at assigning a Hilbert space to $(X,\omega)$ would be to 
set $\mathcal{H}(X)$ equal to the $L^2$--completion of the space of smooth complex--valued 
functions on $X$ with bounded norm with respect to the with inner product
\bea
\langle f_1,f_2\rangle = \int_X (f_1,f_2)\frac{\omega^n}{n!}~.\eea
Notice this inner product includes the K\"ahler potential factor $\exp(-r^2/2)$. This has the added bonus of 
making the measure more convergent. However, 
as is well--known, this Hilbert space is too big -- roughly, in quantum mechanics, the 
wave functions should depend on only ``half'' the phase space variables, for example either 
position or momentum variables. In forming $\mathcal{H}(X)$ we have so far made no such distinction.

The most natural way of solving this problem in general, geometrically, is 
to pick a \emph{polarisation} of $(X,\omega)$. The reader is referred to reference \cite{Woodhouse} for 
the general set--up. Here we note that, since $X$ has a natural complex structure,
namely that induced by the Sasakian structure on $L$, there exists a natural choice of 
complex polarisation, namely the integral distribution $F$ in $T_{\C}X$ spanned by the anti--holomorphic 
vector fields on $X$, $\partial/\partial \bar{z}_i$. The K\"ahler form $\omega$ of course vanishes on 
$F$, and by construction so does the K\"ahler 1--form $\nu$ in (\ref{oneform}). One then 
says that this choice of $\nu$ is \emph{adapted} to the polarisation. The Hilbert space 
$\mathcal{H}$ is then the subspace of polarised elements of $\mathcal{H}(X)$:
\bea
\nabla_X f = 0\eea
where $X\in F$. In the present set--up, with $\nu$ adapted to the polarisation, this reduces to 
\bea
\bar{\partial} f = 0\eea
so that polarised sections of $\mathcal{L}$ may be identified with $L^2$--normalisable 
\emph{holomorphic functions} on $X$. In fact, this Hilbert space, with norm (\ref{inner}), was constructed by 
Berezin in his quantisation of K\"ahler manifolds, introduced in \cite{berezin}.

\subsection{Operators}

The main point of the above construction, however, is that there is a natural map from a certain class of functions on 
$X$ to symmetric operators acting on $\mathcal{H}$, that satisfies Dirac's requirements. Namely,
\bea
A\mapsto\hat{A} \equiv -i\nabla_{X_A} + A = -i\mathcal{L}_{X_A} - (X_A\lrcorner\nu) + A~.\eea
Clearly this is linear, and for any three functions satisfying the Poisson bracket relation $A = [B,C]$, 
the reader can easily check that indeed $-i\hat{A} = [\hat{B},\hat{C}]$. The space of 
\emph{quantisable} functions is then the space of $A$s such that $\hat{A}:\mathcal{H}\rightarrow\mathcal{H}$. 
In particular, $X_A$ must preserve the polarisation $F$, which means that $[X_A,X]\in F$, $\forall X\in F$.

We may now apply this to some of the observables of interest. The Hamiltonian is $H_{\mathrm{BPS}}=r^2/2l$. The 
corresponding Hamiltonian vector field is, up to a factor of $1/l$, just the Reeb vector field $\xi$. This
vector field is holomorphic, and thus preserves the polarisation $F$. Thus the Hamiltonian is 
indeed quantisable\footnote{In general 
this is not guaranteed to be the case, which is one of the  problems with geometric quantisation.}
\bea\label{arse}
lH_{\mathrm{BPS}}\mapsto l\hat{H} = -i\mathcal{L}_{\xi} - (\xi\lrcorner\nu) + r^2/2~.\eea
We must now recall that
\bea
\nu = -\frac{1}{2}i\partial r^2 = -\frac{1}{4}i(\diff+i\diff^c)r^2\eea
where as usual $\diff^c = \mathcal{J}\circ \diff$ and we have
\bea
\partial & = & \frac{1}{2}(\diff+i\diff^c)\nonumber\\
\bar{\partial} & = & \frac{1}{2}(\diff-i\diff^c)~.\eea
Recall also (see {\it e.g.} \cite{MSY2}) that $\diff^c r^2 = 2r^2\eta$, and $\xi\lrcorner\eta=1$, 
$\xi\lrcorner\diff r=0$. Putting all this together, we see that the last two terms in 
(\ref{arse}) cancel, giving
\bea
H_{\mathrm{BPS}}\mapsto\hat{H} = -il^{-1}\mathcal{L}_{\xi}~.
\label{quantumami}
\eea
The energy eigenstates of $\hat{H}$ acting on $\mathcal{H}$ are therefore simply 
holomorphic functions of fixed charge under the Reeb vector field. 

The above calculation in fact generalises to any holomorphic vector field on $X$
that acts isometrically on $(L,g_L)$. Any such vector field $V$ is tangent to the link, meaning that
\bea
\mathcal{L}_V r=0~.\eea
The corresponding Hamiltonian function $A_V$ by definition satisfies
\bea
\diff A_V = -V\lrcorner\omega~.\eea
The unique homogeneous solution to this is
\bea
A_V = \frac{1}{2}r^2\eta(V)~.\eea
It follows that
\bea
-(V\lrcorner\nu) + A_V = 0\eea
where we have used the above equations, together with $\diff r(V)=\mathcal{L}_V r=0$ and the fact that $V$ 
is holomorphic.

A holomorphic Killing vector field $V$ gives rise to a 
conserved quantity in the dynamics of BPS dual giant gravitons. A coordinate along the orbits of $V$ is then canonically conjugate 
to the function $A_V$ on the BPS phase space. On quantisation, this maps to the operator
\bea
\hat{A}_V = -i\mathcal{L}_V\eea
acting on the Hilbert space $\mathcal{H}$ of holomorphic functions. 

In particular, we may apply this to the generators of the $\T^3$ isometry for toric geometries. 
Here one may take $V=\partial/\partial\phi_i$ for any $i=1,2,3$. These give rise to the conserved 
quantities $P_{\phi_i}$, which may be identifed with the symplectic coordinates $y^i$ for 
BPS solutions. On quantisation, these map to operators that we shall call 
$\hat{P}^i = -i\mathcal{L}_{\partial/\partial\phi_i}$, acting on holomorphic functions on $X$.
Applying this to the toric BPS Hamiltonian (\ref{toricBPS}), we recover the result (\ref{quantumami})
\bea
H_{\mathrm{BPS}} \mapsto l^{-1} b_i \hat{P}^i = - il^{-1}{\cal L}_\xi~.
\eea 

Recall that holomorphic functions on $X$ are spanned by elements of the abelian semi--group
\bea
\mathcal{S}_C = \Z^3\cap\mathcal{C}^*\subset \R^3
\eea
of integral points inside the polyhedral cone $\mathcal{C}^*$. We can give a physical interpretation 
of this by using the fact that, upon quantisation, the linear functions $l^a(\mathbf{y})$ map to operators
\bea
l^a(\mathbf{y}) \mapsto  \hat{L}^a = v^a_i \hat{P}^i~.
\eea
Thus, the equations defining $\mathcal{C}^*$ become conditions to be imposed on  states 
$f$ of the Hilbert space
${\cal H}$:
\bea
l^a(\mathbf{y}) \geq 0 \mapsto \langle f,\hat{L}^a f \rangle \geq 0~.
\eea 
The latter precisely means that the quantum numbers of a state $f$, 
which are the eigenvalues of
 $\hat{P}^i$, are $\mathbf{m}\in \mathcal{S}_C $.

\section{Partition functions and entropy minimisation}
\label{partition}

In this section we analyse the classical and quantum partition functions for a BPS dual giant graviton. 
This gives a physical interpretation to the results in \cite{MSY2}. Moreover, we show that 
the classical entropy, viewed as a function of the background Sasakian metric, is minimised for Sasaki--Einstein 
backgrounds. This is rather similar to Sen's entropy function for black holes \cite{sen}, which is also defined 
off--shell and is extremised on solutions. In both cases these extremisation problems allow one to 
compute the entropy of a solution, without knowing its explicit form, but assuming that the solution in fact 
exists. For black holes the extremal entropy typically depends only on the conserved electric and 
magnetic charges, whereas 
the results of \cite{MSY2} determine the extremal entropy of the BPS dual giant in terms of 
equivariant holomorphic invariants of 
the geometry, namely certain Chern classes and weights. For toric geometries, these can be replaced by the toric 
data defining the Calabi--Yau singularity \cite{MSY}.

\subsection{Classical and quantum partition functions}

Given a classical phase space $(X,\omega)$ with Hamiltonian $H$, together with a quantisation
with Hilbert space $\mathcal{H}$ and quantised Hamiltonian $\hat{H}$, one can define   
classical and quantum partition functions. 
The classical partition function is obtained by integrating $\exp(-\beta H)$ over the phase space:
\bea\label{arses}
Z_{\mathrm{classical}}(\beta) = \int_X \e^{-\beta H}\frac{\omega^3}{3!}~,\eea
where $\beta=1/T$ is the inverse temperature, in units where $k_B=1$.
Since we have $H_{\mathrm{BPS}}=r^2/2l$, a change of variable shows that 
(\ref{arses}) is given by
\bea\label{pussy}
Z_{\mathrm{classical}}(\beta) = \frac{l^3}{\beta^3}\int_X \e^{-r^2/2}\frac{\omega^3}{3!}~.\eea
The integrand in this last expression is then easily written in polar coordinates. A simple 
calculation gives
\bea
Z_{\mathrm{classical}}(\beta) = \frac{8l^3}{\beta^3}\vol[L]\eea
where $\vol[L]$ is the volume of the link $(L,g_L)$ \cite{MSY2}. In fact the formula (\ref{pussy}) 
for the volume of $(L,g_L)$ was crucial in \cite{MSY2}. By writing the volume of the link 
in terms of a classical partition function, one can make contact with the formula of 
Duistermaat and Heckman \cite{DH}. This localises the integral on the fixed point set of 
$\xi$, being the Hamiltonian vector field for the Hamiltonian $r^2/2$. Since $\xi$ vanishes 
only at $r=0$, the integral effectively localises at the tip of the cone. One gets a useful 
formula only by taking an appropriate (partial) resolution of the cone $X$. Any such resolution will suffice, and 
the localisation formula expresses the volume, and hence the classical partition function, 
in terms of certain equivariant holomorphic (topological) invariants.
The reader is referred to \cite{MSY2} for further details, which also contains a number of detailed examples.

The quantum (canonical) partition function is equally simple to define. 
This time one takes a trace of the operator $\exp(-\beta\hat{H})$ over the Hilbert space, rather than 
integrating over the classical phase space:
\bea
Z_{\mathrm{quantum}}(\beta) = \mathrm{Tr}_{\mathcal{H}}\,\e^{-\beta\hat{H}}~.\eea
Recall that in section \ref{quantum}, we showed that 
\bea
\beta\hat{H} = \frac{\beta}{l}\mathcal{L}_{r\partial/\partial r}
\eea
when acting on holomorphic functions. Thus we see that the quantum partition function is
precisely\footnote{This is not quite obvious, since in \cite{GMSY} the 
partition function was defined as a trace over all holomorphic functions on $X$, whereas here the Hilbert space 
$\mathcal{H}$ is the space of bounded holomorphic functions, with inner product (\ref{inner}). The traces 
are nevertheless equal. To see this, let $f$ be a holomorphic function with eigenvalue $\lambda> 0$ 
under $l\hat{H}$. Then $f=r^{\lambda}\tilde{f}$ where $\tilde{f}$ is a function on $L$. The $r$ integral in 
the square norm $\|f\|^2$ of $f$ is then finite, since the exponential dominates any monomial in $r$. The 
remaining integral over $L$ is then bounded, since any continuous function on a compact space  is bounded.
} what was defined as the \emph{holomorphic partition function} in \cite{GMSY}, and is essentially the character introduced 
in \cite{MSY2}. 
The variables are related by $t=\beta/l$. Holomorphic functions on $X$ of 
charge $\lambda$ under $\xi$, or equivalently degree $\lambda$ under $r\partial/\partial r$, 
give rise to eigenfunctions of the scalar Laplacian $\nabla^2_L$ on the link $L$. 
Writing
\bea
f=r^{\lambda}\tilde{f}\eea
we have
\bea
-\nabla^2_L \tilde{f} = \lambda(\lambda+4)\tilde{f}~.
\eea
Thus the quantum partition function is a holomorphic analogue of the usual 
partition function of a Riemannian manifold $(L,g_L)$, arising from the spectrum of the 
scalar Laplacian. 

As discussed in \cite{GMSY}, the relation to the character\footnote{See 
\cite{Butti:2006nk} for computations of the character in 
a large number of non--toric examples.} of \cite{MSY2} can be seen as follows.
Given a  holomorphic $(\C^*)^s$ action on $X$, we may define the 
character\footnote{In the following, as in \cite{MSY2}, we will not be 
concerned where sums such as this converge. Similar remarks apply later 
to the grand canonical partition function.}
\be\label{char}
C({\bf q},X) = \mathrm{Tr} \ {\bf q}
\ee
as the trace of the action of 
${\bf q}\in (\C^*)^s$ on the holomorphic 
functions on $X$. Holomorphic functions on $X$ that are 
eigenvectors of the induced $(\C^*)^s$ action 
with eigenvalue $\mathbf{q}^{\mathbf{m}}= \prod_{i=1}^s q_i^{m_i}$ 
form a vector space over $\C$ 
of dimension $d_{\mathbf{m}}$. Each eigenvalue then contributes\footnote{Note that we
include the contribution $\mathbf{m}=\mathbf{0}$, $d_\mathbf{0}=1$, coming from the constant functions.} 
$d_{\mathbf{m}} \mathbf{q}^{\mathbf{m}}$ to the trace (\ref{char}):
\bea
C(\mathbf{q},X) & = & \sum_{\mathbf{m}} d_{\mathbf{m}} \mathbf{q}^{\mathbf{m}}~.
\eea
Letting $\zeta^i$ be a basis for 
the Lie algebra of $U(1)^s\subset (\C^*)^s$, and writing the Reeb vector field
as
\be
\xi = \sum_{i=1}^s b_i \zeta^i~,
\ee
the eigenvalue of $f$ may be written as 
\bea
\lambda_{\mathbf{m}}  = \sum_{i=1}^s b_i \, m_i ~, 
\eea
thus\footnote{{\it cf} the previous footnote. Since, by assumption, $\xi$ 
is in the Lie algebra of $U(1)^s$, any holomorphic function contributing to the trace in the character in \cite{MSY2}
will have bounded norm.}
\bea
\mathrm{Tr}_{\mathcal{H}}\,\e^{-\beta\hat{H}} =  
 \sum_{\mathbf{m}} d_{\mathbf{m}} \,\e^{-\beta\lambda_{\mathbf{m}}/l} =
C(\exp(-\beta \mathbf{b}/l),X) ~.
\eea

It is well known that the volume of $(L,g_L)$ 
arises as the coefficient of a pole in the partition function, which is 
also the trace of the heat kernel \cite{spelling}. One of the  results of 
\cite{MSY2} may then be considered the holomorphic Sasakian analogue of this, namely that
\bea
Z_{\mathrm{quantum}}(\beta) = \frac{1}{(2\pi)^3}Z_{\mathrm{classical}}(\beta)+\mathcal{O}(1/\beta^2)
\eea
as $\beta\rightarrow 0$. This means that the classical and quantum partition functions 
coincide (\ref{agree}) to leading order in $\beta$ as $\beta\rightarrow 0$, 
a familiar result in statistical mechanics.

\subsection{Entropy minimisation}

The classical entropy associated to a single BPS dual giant graviton is given by the standard formula
\bea
S = \frac{\partial}{\partial T}\left(T\log Z_{\mathrm{classical}}\right)\eea
where recall $T=1/\beta$ is the equilibrium temperature. Since 
$Z_{\mathrm{classical}}$ is homogeneous degree three in $T$, we immediately 
deduce that
\bea
\exp (S)= \e^3T^3Z_{\mathrm{classical}}(T=1) = (2l\e T )^3\vol[L,g_L]~,\eea
where we regard $S$ as a function of the Reeb vector field.
The main result of \cite{MSY2} was that Sasaki--Einstein metrics minimise 
the volume in the space of Sasakian metrics on $L$ satisfying
\bea
\mathcal{L}_{\xi}\Omega = 3i\Omega\eea
where $\Omega$ is the (fixed) holomorphic $(3,0)$--form on $X$. This is dual to 
fixing the R--charge of the superpotential in the dual superconformal field theory to be 2.
Thus the entropy function of BPS dual giant gravitons is minimised on backgrounds 
that satisfy the type IIB equations of motion.

\section{Counting BPS states}
\label{grand}

In this final section we introduce a grand canonical partition function that counts 
multiple BPS dual giant gravitons. This may be used to  count 
scalar chiral primaries of the dual superconformal field theory,
and is related to the character of \cite{MSY2}.

\subsection{Grand canonical partition function}

A single BPS dual giant graviton has Hilbert space $\mathcal{H}$. A wavefunction 
is described by a holomorphic function on the K\"ahler cone $X$. It is then straightforward to 
consider states consisting of $n$ BPS dual giant gravitons. Since these are mutually BPS, 
the total energy is just the sum of the individual energies. These multi states are thus 
effectively described by $n$ indistinguishable particles, each with Hilbert space $\mathcal{H}$. 
The $n$--particle Hilbert space 
is hence the symmetric tensor product $\mathcal {H}_n=$ Sym$^n \mathcal{H}$. 

Suppose now that $(X,g_X)$ admits some number $s$ of commuting holomorphic Killing vector 
fields, generated by vector fields $J_i$, $i=1,\ldots,s$. 
Of course, all geometries 
admit at least the Reeb vector field $\xi$ as such a symmetry. We may assume these generate
 a $U(1)^s$ isometry. 
As discussed in section 3, these symmetries
give rise to an $s$--tuple of commuting operators
\bea
\mathbf{\hat{P}}=-i\mathcal{L}_{\mathbf{J}}\eea
acting on $\mathcal{H}$ which quantise the Hamiltonian functions canonically conjugate to the 
corresponding cyclic coordinates in the D3--brane dynamics. It is then natural to construct a
grand canonical partition function that counts multiple BPS dual giant gravitons, weighted by 
these charges, as the trace of $\mathbf{q}^{\hat {\mathbf{P}}}$ over the multi--particle
 Hilbert space\footnote{We define $\mathcal{H}_0=\{1\}$.}
\bea
\mathcal{H}_{\mathrm{multi}} = \bigoplus_{n=0}^\infty \mathcal{H}_n~,
\eea
namely:
\bea\label{multiZ}
{\cal Z}(\zeta,\mathbf{q},X) = \mathrm{Tr}_{\mathcal{H}_{\mathrm{multi}}} 
\zeta^{\hat{N}}\mathbf{q}^{\mathbf{\hat{P}}}~,
\eea
where
\bea
\mathbf{q}^{\mathbf{\hat{P}}} \equiv \prod_{i=1}^s q_i^{\hat{P}^i}~.
\eea
Here $\hat{N}$ is the operator that counts the number of giant gravitons. 
Thus if $\Psi\in \mathcal{H}_n$  one has $\hat{N}\Psi=n\Psi$. 

The usual expression for the grand canonical partition function of a system of indistinguishable
bosonic particles is given by 
\bea
{\cal Z} (\zeta,\mathbf{q},X) = \prod_{\mathbf{m}} \frac{1}{(1-\zeta \mathbf{q}^\mathbf{m})^{d_\mathbf{m}}}
\label{newdef}
\eea
where the product is taken over all states with quantum numbers  $\mathbf{m}$, and 
 $d_\mathbf{m}$ is the degeneracy of states with equal quantum numbers\footnote{Notice
 that the infinite product includes the term with $\mathbf{m}=\mathbf{0}$, $d_\mathbf{0}=1$.}. 
It is then easy to express ${\cal Z} (\zeta,\mathbf{q},X)$ in terms of the 
character $C(\mathbf{q},X)$.
By taking the logarithm of (\ref{newdef}) and expanding the terms 
$\log (1-\zeta \mathbf{q}^\mathbf{m})$ in a formal power series, we obtain 
\bea
{\cal Z} (\zeta,\mathbf{q},X) = \exp\left[ \sum_{n=1}^\infty \frac{\zeta^n}{n}\, C(\mathbf{q}^n,X)\right]~.
\label{trivial}
\eea
One may then formally expand
\bea
{\cal Z} (\zeta,\mathbf{q},X) = \sum_{n=0}^{\infty} \zeta^n Z_n(\mathbf{q},X)
\label{formal}
\eea
where $Z_n(\mathbf{q},X)$ counts the $n$--particle states. 
In particular, notice that the single--particle partition function 
\bea
Z_1(\mathbf{q},X) = C(\mathbf{q},X)\eea
is precisely the character.

The argument of \cite{nemani,smith}, showing that 
the number of BPS dual giants is bounded from above by $N$ (the number of 
background D3--branes), applies, since it is entirely based on considerations in AdS$_5$.
Thus, the physical quantity of interest is a truncation of (\ref{formal}) to order $N$.

\subsubsection*{Toric geometries}

In the case $(L,g_L)$ is toric,  the K\"ahler cone $X$ is an affine toric variety. 
The Hamiltonian $\T^3$ action 
fibres $X$ over a conical convex subspace $\mathcal{C}^*$ of $\R^3$, which is the image 
of the moment map
(\ref{momentmap}).
Equivalently, we may specify the abelian semi--group 
\bea
\mathcal{S}_C = \Z^3\cap\mathcal{C}^*\subset \R^3~.\eea
It is a standard result that holomorphic functions on $X$ are spanned by
 elements of the 
semi--group $\mathbf{m}\in\mathcal{S}_C$, as we already discussed earlier. 
An $n$--particle BPS dual giant state, which is an 
eigenstate under the torus action, is then a vector
\bea
|\mathbf{m_1},\mathbf{m_2},\ldots,\mathbf{m_n} \rangle \, \in \mathrm{Sym}^n \mathcal{S}_C~,\eea
and the grand canonical  partition function is hence 
\bea\label{toricgrand}
{\cal Z} (\zeta,\mathbf{q},X) = \prod_{\mathbf{m}\,\in \mathcal{S}_C} 
\frac{1}{1-\zeta \mathbf{q}^{\mathbf{m}}}=
1+ \zeta  \sum_{\mathbf{m}\in \mathcal{S}_C}\mathbf{q}^\mathbf{m} + {\cal O}(\zeta^2) ~.
\eea
Notice that in this case there is precisely one holomorphic function for each set of quantum numbers 
$\mathbf{m}\in \mathcal{S}_C$, thus $d_\mathbf{m}=1$.

The simplest example is that of $S^5$,  discussed in subsection \ref{toricsub}. 
The K\"ahler cone over the round $S^5$ is of course $\C^3$, with its standard K\"ahler form $\omega$. 
The Cartan subgroup $U(1)^3\subset U(3)$ preserves $\omega$ and the corresponding moment map 
(\ref{momentmap}) has image $\mu(\C^3)=(\R_+)^3$. The semi--group for this affine toric variety 
is thus
\bea
\mathcal{S}_C = \Z^3\cap(\R_+)^3 = (\Z_+)^3 = \{\mathbf{m}=(m_1,m_2,m_3)\mid m_i\geq 0, i=1,2,3\}~.\eea
Equivalently, recalling  that $v^a_i=\delta^a_i$, the fact that $m_i\geq 0$ follows from the 
argument at the end of section \ref{quantum}.
Thus the Hilbert space is isomorphic to that of the three--dimensional 
harmonic oscillator, with $l\hat{H}$ being the energy operator\footnote{Note 
that the ground state energy is zero.}, 
and the partition function (\ref{toricgrand}) precisely reduces to that in \cite{giants}.
Alternatively, it is given by (\ref{trivial}), with 
\bea
C(\mathbf{q}^n,\C^3) = \frac{1}{(1-q_1^n)(1-q_2^n)(1-q_3^n)}~.
\eea

\subsection{Counting BPS operators in the dual SCFT}

According to the AdS/CFT dictionary, BPS states in the geometry are dual to BPS operators
in the SCFT, with the same quantum numbers. It is then natural to interpret the dual giant gravitons
that we have considered in terms of BPS operators of the dual conformal field theory.

Let us first recall how this correspondence works in the prototypical example of
$\mathcal{N}=4$ super Yang--Mills. 
Generic  (1/8--BPS) single--trace scalar chiral primary operators of the ${\cal N}=4$ super 
Yang--Mills theory are of the type 
\bea
\mathrm{Tr} ({X}^{m_1}{Y}^{m_2}{Z}^{m_3}) \quad \qquad 
\Delta = m_1 + m_2 + m_3 =\sum_{i=1}^3 m_i b_i\label{oper}
\eea 
where $X,Y,Z$ are the three complex scalar fields, in the adjoint of $U(N)$. 
In the abelian theory, with $N=1$, these operators are simply
 monomials in three complex variables $x,y,z$, 
of the type
$x^{m_1}y^{m_2}z^{m_3}$ with $m_i\geq 0$, and  span the coordinate 
ring of  $\C^3$. When $N > 1$, and including also multi--trace operators,
one obtains monomials in the eigenvalues of the three operators, which span the coordinate
 ring of the symmetric product $\mathrm{Sym}^N \C^3$. In other words, the scalar sector of
the chiral ring of the ${\cal N}=4$ super Yang--Mills theory, for finite $N$, 
 is isomorphic to the ring of holomorphic
functions on $\mathrm{Sym}^N \C^3$ (for more details see, {\it e.g.} \cite{emergent}).
The results of \cite{giants} then show that 
this space arises from  quantising the phase spaces of precisely $N$ non--spinning BPS
dual giant gravitons in AdS$_5\times S^5$. The reason why one considers the Hilbert space for precisely $N$ dual giants is that these are viewed as 
excitations of the background $N$ D3--branes.
The $N$--particle Hilbert space is the symmetric tensor product 
$\mathcal{H}_N=$ Sym$^N \mathcal{H}$ of the single--particle Hilbert space $\mathcal{H}$, and  
the partition function for $N$ BPS dual giant gravitons may be obtained as the  
coefficient $Z_N(\mathbf{q},\C^3)$ of $\zeta^N$ in (\ref{multiZ}). In fact, this partition function 
agrees precisely with the counting of mesonic scalar\footnote{The full chiral ring in general contains
operators with non--zero spin. These are accounted for in the index of \cite{Kinney:2005ej}. However,
throughout 
this paper, we restrict our attention to spinless configurations.} chiral primary operators in the gauge theory, 
obtained using the index of \cite{Kinney:2005ej}.

More generally, if a SCFT arises from the IR limit of $N$ D3--branes 
at an isolated singularity $X$, then
the classical vacuum 
moduli space should be the symmetric product $\mathrm{Sym}^N X$. 
The Hilbert space of $N$ BPS dual giants  is then spanned 
by the same set of generators (as a $\C$--algebra) as the 
ring of holomorphic functions on $\mathrm{Sym}^N X$. This is the scalar sector of the chiral ring 
of the dual superconformal field theory.
When the singularity $X$ admits a crepant resolution\footnote{We 
note that there are plenty of examples of $X$, admitting 
Ricci--flat K\"ahler cone metrics, which have no crepant resolution. 
In this case, the dual SCFTs might in principle 
be quite exotic, and in particular 
not be described by quiver gauge theories.}, one 
expects to be able to describe the dual SCFT by a quiver gauge theory. 
In this case, the mesonic scalar chiral primary operators are 
constructed from closed loops in the quiver, by matrix--multiplying the 
corresponding bifundamental fields. 
These gauge--invariant operators then generate the 
ring of holomorphic functions on the vacuum moduli space. Thus 
we see that the partition function $Z_N(\mathbf{q},X)$  for $N$ dual giants
also counts the mesonic scalar chiral primary operators in the dual SCFT, 
weighted by their $U(1)^s$ flavour charges.
This is in full agreement with the results of \cite{Kinney:2005ej}.
We emphasize that this partition function 
is related simply to the character $C(\mathbf{q},X)$, and the latter 
may be computed using only 
a minimal amount of geometric information. In particular, one 
may apply localisation techniques to compute $C(\mathbf{q},X)$, as described in \cite{MSY2}.

\vskip 1cm

\noindent{\bf Note}: Just as this paper was about to be submitted to the arXiv, we became aware 
of \cite{Benvenuti:2006qr}. Their results overlap with our section \ref{grand}.
Their conclusions are in agreement with ours.

\subsection*{Acknowledgments}
\noindent 
We thank C. Beasley, J. F. Morales, and S.--T. Yau for discussions,  
and S. Minwalla for communications. J. F. S. would like to thank 
the organisers of the Fourth Simons Workshop in Mathematics and Physics 
for hospitality during the final stages of this work. Some of the 
results of this paper were mentioned in a seminar at that workshop.
J. F. S. is supported by NSF grants DMS--0244464, DMS--0074329 and DMS--9803347.

\appendix

\section{$\kappa$--symmetry analysis}
\label{BPS}

In this appendix we demonstrate that the dual giant graviton solutions considered in the main text
are precisely the set of BPS solutions. That is, they are the general set of solutions of the D3--brane dynamics respecting 
$\kappa$--symmetry of the world--volume  action. One must thus impose that the Killing spinor $\epsilon$ 
of type IIB supergravity, 
\bea
\nabla_M \epsilon +\frac{i}{192} F_{MP_1P_2P_3 P_4}\Gamma^{P_1P_2P_3P_4}\epsilon = 0~,\label{labe}
\eea
in the background of AdS$_5\times L$, also satisfies the $\kappa$--symmetry projection 
\bea
\Gamma_\kappa \epsilon & = & i \epsilon ~,
\label{pkappa}
\eea
where the $\kappa$--symmetry projection matrix is defined as 
\bea
\Gamma_\kappa & = &  \frac{1}{4!\sqrt{- \det G_{\mu\nu}}}
\epsilon^{\mu\nu\rho\sigma} \gamma_{\mu\nu\rho\sigma}~.
\eea
The $\gamma$--matrices above are the world--volume gamma matrices, defined as
\bea
\gamma_{\mu} & = & \frac{\de X^M}{\de \zeta^\mu } e^{\hat{M}}_M \Gamma_{\hat{M}}~,
\eea
where $e^{\hat{M}}_M$ is a vielbein, that is, a local 
orthonormal frame, and $\Gamma_{\hat{M}}$ are ten--dimensional 
flat spacetime gamma matrices, obeying
\bea
\{ \Gamma_{\hat{M}} , \Gamma_{\hat{N}}\} = 2\eta_{\hat{M}\hat{N}}~.
\eea
We find
\bea
&&\gamma_0 = V(R)^{1/2} \Gamma_{\hat{0}} + \frac{\dot R }{V(R)^{1/2}}\Gamma_{\hat{1}} +l \left(
(\dot \psi+ \sigma_{\alpha} \dot x^{\alpha})\Gamma_{\hat{5}}+ \dot x^{\alpha} \,\hat{e}^{\hat{\alpha}}_{\alpha}\,
\Gamma_{\hat{\alpha}+5}\right)\nonumber\\[2mm]
&&\gamma_1 = R \Gamma_{\hat{2}} \quad  \gamma_2 = R\cos\zeta^1 \Gamma_{\hat{3}} \quad 
\gamma_3 = R \sin \zeta^1 \Gamma_{\hat{4}}~,
\eea
where  $\hat{e}^{\hat\alpha}_\alpha  =  {e}^{\hat{\alpha}+5}_{\alpha+5}$, $\alpha=1,\dots,4$, 
 is a vielbein such that 
$h_{\alpha\beta}= \hat{e}^{\hat{\alpha}}_{\alpha}\, \hat{e}^{\hat{\beta}}_{\beta}\,\delta_{\hat\alpha\hat\beta}$.
The $\kappa$--symmetry projector is then 
\bea
\Gamma_\kappa  =  \Delta^{-1/2}\left[V(R)^{1/2} \Gamma_{\hat{0}} + \frac{\dot R }{V(R)^{1/2}}\Gamma_{\hat{1}} +l \left(
(\dot \psi+ \sigma_{\alpha} \dot x^{\alpha})\Gamma_{\hat{5}}+ \dot x^{\alpha} \,\hat{e}^{\hat{\alpha}}_{\alpha}\,
\Gamma_{\hat{\alpha}+5}\right) \right] \Gamma_{\hat{2}\hat{3}\hat{4}}~.
\eea

In the following we adopt the conventions of \cite{paper5}.
Using the spinor ansatz
\bea
\epsilon & = & \Psi \otimes \chi \otimes \theta
\eea
where $\epsilon$ is the complexified type IIB spinorial parameter obeying, 
in our conventions, the following chirality projection
\bea
\Gamma_{11}\epsilon = -\epsilon~,
\label{chira}
\eea
(\ref{labe}) reduces to the two equations
\bea
\nabla_m \Psi +\frac{1}{2l}\rho_m\Psi &=& 0~, \qquad m =0,\dots,4\\
\nabla_{\alpha}\chi -\frac{i}{2l}\gamma_{\alpha} \chi & = & 0~,\qquad \alpha=0,\dots,4~,
\eea
where we now take the $\alpha$ index to run from 0 to 4. 
These may be recognised as the standard equations obeyed by any Killing spinor $\Psi$
of AdS$_5$ (of radius $l$) and the Killing spinor $\chi$ of an arbitrary Sasaki--Einstein manifold with 
line element $l^2\diff s^2_L$ and corresponding Ricci tensor $\mathrm{Ric}=4l^2g_L$. 
Notice that, in the notation of \cite{paper5}, $f=-4/l$ is the overall constant factor multiplying the Ramond--Ramond five--form. 
The warp factor has been set to zero.
We have also chosen the following decomposition of the ten--dimensional Dirac matrices:
\bea
\Gamma_{\hat{m}} & = & \rho_{\hat{m}} \otimes \1\otimes \sigma^3~,\qquad \hat{m}=0,\ldots,4\nonumber\\
\Gamma_{\hat{\alpha}+5} & = & \1 \otimes \gamma_{\hat{\alpha}} \otimes \sigma^1~, \qquad \hat{\alpha}=0,\ldots,4~,
\eea
where 
\bea
\{\rho_{\hat{m}},\rho_{\hat{n}}  \} = 2 \eta_{\hat{m}\hat{n}}~, \qquad \{\gamma_{\hat{\alpha}},\gamma_{\hat{\beta}} \} = 2\delta_{\hat{\alpha}\hat{\beta}}~,
\eea
are (flat) gamma matrices of Cliff$(4,1)$ and Cliff$(5,0)$, respectively, and $\sigma^1,\sigma^2,\sigma^3$ are 
the Pauli matrices.

We may now return to the $\kappa$--symmetry projection. First note that 
$\gamma_{\hat{\alpha}}\chi$, $\hat{\alpha}=0,1,\ldots,4$ are linearly independent spinors, over the \emph{real} numbers, on $L$. 
For, consider the linear combination
\bea
\sum_{\hat{\alpha}=0}^4 a_{\hat{\alpha}}\gamma_{\hat{\alpha}}\chi=0\eea
with $a_{\hat{\alpha}}\in\R$. One may now apply $\bar{\chi}\gamma_{\hat{1}}$ on the left to obtain
\bea\label{noddy}
a_{\hat{1}} + a_{\hat{2}}\bar{\chi}\gamma_{\hat{1}}\gamma_{\hat{2}}\chi=0~.\eea
Here we have used $\bar{\chi}\chi=1$, together with the fact that 
\bea
-\frac{i}{2}\bar{\chi}\gamma_{\alpha\beta}\chi\diff x^{\alpha}\wedge\diff x^{\beta} = \omega_T = \hat{e}^{\hat{1}} \wedge \hat{e}^{\hat{2}} + \hat{e}^{\hat{3}} \wedge \hat{e}^{\hat{4}}
\eea
is the transverse K\"ahler form. Also note that the Killing spinor $\chi$ obeys the projection
 (see {\it e.g.} \cite{Gary})
\bea
\gamma_{\hat{0}} \chi & = & \pm \chi~,
\label{sean}
\eea
and $\bar{\chi}\gamma_{\hat{\alpha}}\chi =0$ for $\hat{\alpha}=1,\ldots,4$. In particular, the second term in (\ref{noddy}) is pure imaginary, 
and this immediately gives $a_{\hat{1}}=a_{\hat{2}}=0$. A similar argument gives $a_{\hat{3}}=a_{\hat{4}}=0$ and thus we also have $a_{\hat{0}}=0$.
Using these facts we see that the $\kappa$--symmetry condition (\ref{pkappa}) implies that
\bea
\dot{x}^{\alpha}=0~, \qquad \alpha=1,\ldots,4~.
\eea

Also note that, since $\psi$ is a cyclic coordinate ($\de / \de
\psi $ is a Killing vector field), the conjugate momentum is 
\bea
P_\psi= k~,
\eea
a constant. Using the expression for $P_{\psi}$ in terms of $\dot \psi$, we have 
\bea
\dot \psi = \frac{l^2k\Delta^{1/2}}{NR^3}
\eea
with 
\bea
\Delta = \frac{V(R)-\frac{\dot R^2}{V(R)}}{1+\frac{l^6k^2}{N^2R^6}}~.
\eea
Inserting this into  $\Gamma_\kappa$ we obtain 
\bea
\Gamma_\kappa & = & \Delta^{-1/2}\left[V(R)^{1/2}\Gamma_{\hat{0}} +\frac{\dot{R}}{V(R)^{1/2}}
\Gamma_{\hat{1}} + \frac{l^3k\Delta^{1/2}}{NR^3}\Gamma_{\hat{5}} \right] \Gamma_{\hat{2}\hat{3}\hat{4}}~.
\eea
It is convenient to define
\bea
\frac{R}{l} & \equiv & \sinh \alpha~,
\eea
so that the projection (\ref{pkappa}) becomes
\bea
&&\left[\cosh \alpha \,\rho_{\hat{0}} \otimes \1 \otimes \1 +\frac{\dot{R}}{\cosh\alpha}\rho_{\hat{1}}\otimes \1
 \otimes \1 
 + \frac{l^3k\Delta^{1/2}}{NR^3} \1 \otimes \gamma_{\hat{0}} \otimes 
i \sigma^2 \right] \epsilon = \nonumber\\ 
& &=
\Delta^{1/2}\, i 
\rho_{\hat{2}\hat{3}\hat{4}}\otimes \1 \otimes \1 \, \epsilon~.
\label{prot}
\eea

Finally, imposing the chirality condition (\ref{chira}), $\sigma^2 \theta = - \theta$, 
after a little algebra, the projection (\ref{prot}) reduces to 
\bea
\Lambda\Psi\equiv\left[\cosh \alpha \, \1 - \frac{\dot{R}}{\cosh\alpha}\rho_{\hat{0}}\rho_{\hat{1}}+\Delta^{1/2} \rho_{\hat{1}} 
\pm  \frac{l^3k\Delta^{1/2}}{NR^3}   i \rho_{\hat{0}} \right]\Psi & = & 0~.
\eea

To proceed, we shall need an explicit form of the Killing spinor in AdS space.
A useful expression, that may be adapted
for instance from \cite{giants}, reads, in our notation:
\bea
\Psi = \e^{-\frac{\alpha}{2}\rho_{\hat{1}}}\, \e^{-\frac{t}{2l}\rho_{\hat{0}}}\,
\e^{\frac{\theta}{2}\rho_{\hat{1}\hat{2}}}
\,\e^{\frac{\phi_1}{2}\rho_{\hat{1}\hat{3}}}\,\e^{\frac{\phi_2}{2}\rho_{\hat{2}\hat{4}}} \Psi_0 
\equiv D \Psi_0~,
\eea
where $\Psi_0$ is a constant spinor. In order to compute $\Lambda$ acting on $\Psi_0$
we need to commute this through $D$. It is useful to record the following identities:
\bea
&&\rho_{\hat{0}} D = D \left[ \cosh \alpha \, \1 + \sinh\alpha\, \e^{\frac{t}{l}\rho_{\hat{0}}}\left[\cos\theta\, \e^{-\phi_1
\rho_{\hat{1}\hat{3}}}\rho_{\hat{1}} +\sin \theta\, \e^{-\phi_2 \rho_{\hat{2}\hat{4}}}\rho_{\hat{2}} \right] \right]\rho_{\hat{0}}\equiv D A~,\nonumber\\
&&\rho_{\hat{1}} D = D \,\e^{\frac{t}{l}\rho_{\hat{0}}}\left[\cos\theta\, \e^{-\phi_1
\rho_{\hat{1}\hat{3}}}\rho_{\hat{1}} +\sin \theta\, \e^{-\phi_2 \rho_{\hat{2}\hat{4}}}\rho_{\hat{2}} \right] \equiv D B~.
\eea
We then find 
\bea
\Lambda D \Psi_0 = D \left[\cosh \alpha \, \1 - \frac{\dot R}{\cosh \alpha} AB +\Delta^{1/2} B \pm
i \frac{l^3k\Delta^{1/2}}{NR^3} A \right] \Psi_0 = 0~. 
\label{abba}
\eea

First, let us restrict to the solutions (\ref{BPSsolu}) and check that they are indeed BPS.
Thus, let us set
\bea
\dot R =0 \qquad k= \frac{NR^2}{l^2} \qquad \Delta^{1/2} = \frac{R_0}{l}= \sinh \alpha_0~.
\eea
Equation (\ref{abba}) reduces to 
\bea
D\left[\cosh \alpha_0 \, \1 +  \sinh\alpha_0\, \e^{\frac{t}{l}\rho_{\hat{0}}}\left[\cos\theta\, \e^{-\phi_1
\rho_{\hat{1}\hat{3}}}\rho_{\hat{1}} +\sin \theta\, \e^{-\phi_2 \rho_{\hat{2}\hat{4}}}\rho_{\hat{2}} \right]\right] 
\left(\1 \pm i \rho_{\hat{0}} \right)
\Psi_0 = 0  
\eea
and thus we simply require that 
\bea
i \rho_{\hat{0}}  \Psi_0 = \mp \Psi_0  ~.
\label{require}
\eea
This can always be achieved, as can be seen for instance from an explicit expression for the 
$\rho_{\hat{0}}$ matrix. One can choose the following  basis of Dirac matrices
in AdS$_5$\footnote{The construction is standard, see {\it e.g.} \cite{polchinski}.} 
\bea
\rho_{\hat{0}}  =  \left(
\begin{array}{cc}
0 & 1 \\
-1 & 0
\end{array}
\right)\otimes \1\qquad
\rho_{\hat{1}}  =  \left(
\begin{array}{cc}
0 & 1 \\
1 & 0
\end{array}
\right)\otimes \1\nonumber\\
\rho_{\hat{2}}  = \1 \otimes \left(
\begin{array}{cc}
0 & 1 \\
1 & 0
\end{array} \right)
\qquad ~
\rho_{\hat{3}}  =  \1\otimes \left(
\begin{array}{cc}
0 & -i \\
i & 0
\end{array} \right)
\eea
and $\rho_{\hat{4}} = -i \rho_{\hat{0}\hat{1}\hat{2}\hat{3}}$. We may then take 
\bea
\Psi_0 = \left(\begin{array}{c}i \\ \mp 1\end{array} \right) \otimes \tilde{\Psi}_0
\label{explicit}
\eea
The condition (\ref{require}) is the AdS analogue of (\ref{sean}) for Sasaki--Einstein manifolds.

Finally, let us show that in fact the solutions we considered are the set of all BPS solutions.
The projections must hold at any point of the world--volume of the D3--brane, thus we may simplify
the calculation by conveniently setting $t=\theta=\phi_1=\phi_2=0$. Then
\bea
A & = & \left(\cosh \alpha \, \1 + \sinh\alpha \, \rho_{\hat{1}} \right) \rho_{\hat{0}}\nonumber\\
B & = & \rho_{\hat{1}}~.
\eea
Next, we choose a spinor that obeys the projection (\ref{require}). Thus, in our particular basis, we
may choose (\ref{explicit}). It follows that 
\bea
 \bar \Psi_0  \Psi_0 = 2 \, \|\tilde{\Psi}_0 \|^2\qquad \quad   \bar \Psi_0 \rho_{\hat{1}} \Psi_0 = 0 
\eea
and in particular, applying $\bar \Psi_0 D^{-1}$ to the left of equation (\ref{abba}), we obtain 
\bea
\cosh \alpha \pm i\, \dot R \tanh \alpha - \frac{l^3k\Delta^{1/2}}{NR^3}\cosh\alpha = 0~. 
\eea
Thus we conclude that necessarily 
\bea
\dot R = 0 \qquad k = \frac{R^2N}{l^2}~,
\eea
while the remaining components of (\ref{abba}) proportional to $\rho_{\hat{1}}$ are 
automatically satisfied. This concludes our proof that (\ref{BPSsolu}) are 
all the $\kappa$--symmetric solutions to the D3--brane
equations of motion.


\section{A change of coordinates}

\label{change}

In this appendix we give an explicit change of coordinates between
Sasakian coordinates and symplectic coordinates, in the case that the 
K\"ahler cone $X$ is toric.

When the cone is toric, the metric may be written as either
\bea
\dd s^2_X & = & G_{ij} \dd y^i \dd y^j + G^{ij} \dd\phi_i \dd\phi_j\qquad i,j =1,2,3
\eea
or as
\bea
\dd s^2_X &=& \dd r^2 + r^2(\dd\psi + \sigma)^2 +r^2 h_{\alpha\beta}\dd x^{\alpha} \dd x^{\beta}\qquad \alpha,\beta=1,2,3,4
\eea
where 
\bea
r^2 &=& 2 b_i y^i\nonumber\\
h_{\alpha\beta}\dd x^{\alpha} \dd x^{\beta} &=& H_{pq} \dd \eta^p \dd\eta^q +
 H^{pq} \dd\varphi_p \dd \varphi_q\qquad p,q=1,2\\
\sigma &=& 2 \eta^p \dd\varphi_p\nonumber
\eea
and we have used that, locally, the transverse K\"ahler metric is also toric. Now define 
\bea
\tilde \varphi_i & = & (\psi,\varphi_1,\varphi_2)_i
\eea
and suppose that the angular variables in the above metrics are linearly related
\bea
\phi_i & = & A_{i}^{\ j} \tilde \varphi_j~.
\label{B52}
\eea
This allows one to determine the matrix $G^{ij}$ by direct comparison. Inverting
this, we see that 
\bea
G & = & A M A^T
\eea
with 
\bea
M & = & \frac{1}{r^2}\left(
\begin{array}{cc}
1+ 4 H_{rs}\eta^r \eta^s & -2 H_{rq}\eta^r\\
- 2H_{pr}\eta^r & H_{pq}
\end{array}
\right)~.
\eea
It may then be verified that the following is the change of coordinates for  the non--angular part of the metric
\bea
y^i =  r^2 (A^{-T})^i_{\ j}w^j\qquad w^j = \left(\frac{1}{2},\eta^1,\eta^2\right)^j~.
\eea
There is a consistency condition that the matrix $A$ must satisfy. In
particular,
\bea
A_{i}^{\ 1} y^i& = & \frac{1}{2}r^2
\eea
implies that $A_{i}^{\ 1}=b_i$. Thus we must set
\bea
A & = & \left(
\begin{array}{ccc}
b_1  & * & *\\
b_2  & * & *\\
b_3 & * & *\\
\end{array}
\right)
\eea
with the entries $*$ arbitrary, provided that $A$ is invertible.
Note that the Reeb vector transforms as 
\bea
\frac{\de}{\de \psi} = \frac{\de \phi_i}{\de \psi} \frac{\de}{\de \phi_i }=
b_i \frac{\de}{\de \phi_i}~.
\label{B13}
\eea
Finally, using the fact that all the transverse K\"ahler coordinates are constants
for BPS motion ({\it cf} (\ref{BPSsolu})), we see from (\ref{B52}) that 
\bea
\dot \phi_i & = & b_i \dot \psi ~.
\label{B14}
\eea
This may be used to obtain a quick derivation of the BPS solutions for toric geometries,
discussed in the main text.


\end{document}